\documentstyle[emulateapj]{article}
\def\keywords{}

\def\aap{A\&A}
\def\apjl{ApJ}
\def\apjs{ApJS}

\begin{document}

\def\nth{n_{\rm th}}
\def\nobs{n_{\rm obs}}
\def\dmin{d_{\rm min}}
\def\macho{{\sc macho}}
\def\newpage{\vfill\eject}
\def\vs{\vskip 0.2truein}
\def\gnu{\Gamma_\nu}
\def\fnu {{\cal F_\nu}}
\def\mass{m}
\def\lum{{\cal L}}
\def\imf{\xi(\mass)}
\def\ilf{\psi(M)}
\def\msun{M_\odot}
\def\zsun{Z_\odot}
\def\met{[M/H]}
\def\vi{(V-I)}
\def\mtot{M_{\rm tot}}
\def\mhalo{M_{\rm halo}}
\def\pp{\parshape 2 0.0truecm 16.25truecm 2truecm 14.25truecm}
\def\la{\mathrel{\mathpalette\fun <}}
\def\ga{\mathrel{\mathpalette\fun >}}
\def\fun#1#2{\lower3.6pt\vbox{\baselineskip0pt\lineskip.9pt
  \ialign{$\mathsurround=0pt#1\hfil##\hfil$\crcr#2\crcr\sim\crcr}}}
\def\ie{{ i.e., }}
\def\eg{{ e.g., }}
\def\etal{{et al.\ }}
\def\etalc{{et al., }}
\def\kpc{{\rm kpc}}
 \def\Mpc{{\rm Mpc}}
\def\mh{\mass_{\rm H}}
\def\mmax{\mass_{\rm u}}
\def\ml{\mass_{\rm l}}
\def\bc{f_{\rm cmpct}}
\def\br{f_{\rm rd}}
\def\kmsec{{\rm km/sec}}
\def\ibl{{\cal I}(b,l)}
\def\dmax{d_{\rm max}}
\def\dmin{d_{\rm min}}
\def\mbol{M_{\rm bol}}
\def\kms{{\rm km}\,{\rm s}^{-1}}

\lefthead{GRAFF \& GAUDI}
\righthead{DIRECT DETECTION OF GIANT CLOSE-IN PLANETS}
%%%%%%%%%%%%%%%%%%%%%%%
%%%%%%%electronic submission format
%\submitted{Version of \today}
\title{Direct Detection of Giant Close-In Planets Around the Source Stars of
Caustic-Crossing Microlensing Events}
\author{David S. Graff and B. Scott Gaudi}
\affil{Departments of Astronomy and Physics, The Ohio State University,
Columbus, OH 43210, USA}
\authoremail{graff.25@osu.edu, gaudi@astronomy.ohio-state.edu}
%%%%%%%%%%%%%%%%%%%%%%%%%%%%%%%

\begin{abstract}

We propose a direct method to detect close-in giant planets
orbiting stars in the Galactic bulge.  This method uses
caustic-crossing binary microlensing events discovered by survey teams
monitoring the bulge to measure light from a planet orbiting the
source star.  When the planet crosses the caustic, it is more
magnified than the source star; its light is magnified by two orders
of magnitude for Jupiter size planets.  If the planet is a giant close
to the star, it may be bright enough to make a significant deviation
in the light curve of the star.  Detection of this deviation requires
intensive monitoring of the microlensing light curve using a 10-meter
class telescope for a few hours
after the caustic.  This is the only method yet proposed to
directly detect close-in planets around stars outside the solar
neighborhood. 

\end{abstract}

\keywords{planetary systems --- gravitational lensing}
%%%%%%%%%%%%%%%%%%%%%%%%%%%%

\setcounter{footnote}{0}
\renewcommand{\thefootnote}{\arabic{footnote}}

\section{Introduction}

One of the scientific goals of microlensing searches towards the
Galactic bulge is the detection of planets orbiting the primary
lenses.  These searches are
conducted in the following manner: One of the microlensing searches,
EROS (\cite{erostrigger}) or OGLE (\cite{udalski1994}) or the now
terminated MACHO program (\cite{alcock1996}) launches an electronic
alert of an ongoing microlensing event.  These events are then
monitored by follow-up groups such as the PLANET (\cite{albrow1998}),
MPS (\cite{mps}), or GMAN (\cite{alcock1997}) collaborations.  While
the searching teams typically monitor stars $\sim$ once per day, the
follow-up campaigns, with a network of telescopes around the globe,
sample much more frequently.  A planet orbiting the primary lens 
with semi-major axis $a$ in the
``lensing zone'', $0.6-1.5 R_E$ (\cite{mandp1991}; \cite{gandl1992}),
where the Einstein radius, $R_E$ is
\begin{equation}
R_E=\left ( \frac{4 G M}{c^2}\frac{D_{L}D_{LS}}{D_S} \right )^{1/2} \, ,
\end{equation}
may cause detectable deviations from the standard microlensing light
curve (see \cite{sackett1999} for a review).  Here, $D_L$, $D_S$, and $D_{LS}$ are respectively the distances to the lens, the source star, and between the lens and source stars.
For a bulge source lensed by a 0.3 $\msun$ lens, the lensing zone, where one is most sensitive to
planets, is in the range $1.2 - 3.2$ AU.  

In this paper, we discuss another means by which a microlensing
follow-up experiment may detect a planet, in this case a giant planet
close ($a\la0.1~{\rm AU}$) to the {\it source} star.  Confounding prior expectations, such
close-in planets are found to be relatively common.  They have been detected by
several collaborations (\cite{mandq1995},
\cite{cochranetal1997}, \cite{noyesetal1997}; see \cite{mandb1998} for
a review)
and can be found around $\sim 1\%$ of all stars (\cite{mandb2000}).

Since direct detection is most sensitive to planets close to the
source ($a\la0.1~{\rm AU}$), it is complementary to the 
traditional microlensing light curve deviation
method which can only detect planets in the lensing zone ($a\simeq 1-3~{\rm AU}$).  Unlike
other methods of planet detection, such as radial velocity
measurements or astrometric shifts\footnote{It may also be possible to
indirectly detect planets orbiting stars in the Galactic bulge using occultation.}, this method can be
used to directly detect planets in
the bulge of our Galaxy, thus allowing us to
compare planet formation under conditions different from those in the solar
neighborhood.

\subsection{Caustics}

All strong gravitational lenses produce ``caustics,'' regions in which
a point source is (formally) infinitely magnified.  In reality, no
source is truly point-like, and the passage of the source through the
caustic allows us to spatially resolve the source.  Already,
limb-darkening profiles have been measured in sources as far away as
the Small Magellanic Cloud (\cite{afonso2000}), and it has been
proposed that star-spots could be imaged when a source passes through
a caustic (\cite{hands2000}; \cite{hkcp2000}; \cite{bryce2000}).  In this
paper, we examine the possibilities of directly detecting light from a
planet as it passes through a caustic.

There are several different types of caustics depending on the lens
configuration.  A single point lens has a point caustic corresponding
to perfect alignment between the source and the lens, when the image
of the source is the Einstein ring.  These events are rare since
they require such perfect alignment.  In the case of binary lenses,
the caustics form a network of ``folds'' and ``cusps.''  The caustic
structure of binary lenses is cataloged in Schneider \& Weiss (1986)
and Erdl \& Schneider (1993).

Because caustics form closed curves, a caustic light curve generically
has pairs of crossings.  The time of the second caustic crossing can
often be predicted days in advance, allowing for scheduling of
detailed monitoring of the caustic crossing.

The best studied caustic crossing event, which allows us to illustrate
the technology, is MACHO-98-SMC-1 (\cite{alcock1999}), an event which took place in the Small Magellanic Cloud.  The MACHO
group first sent an alert indicating that a microlensing event was
taking place.  After the first caustic crossing, the MACHO group
launched a second alert with a rough prediction of when the second
caustic crossing would occur.  As the second caustic crossing
approached, predictions of its time became more accurate so that when
it occurred, several groups were able to devote all their telescope
time for one night to the observation of this one event.  Notably, the
passage of the source through the caustic was imaged every $\sim 5$
minutes by the PLANET collaboration (\cite{albrow1999}) from South
Africa, followed immediately, and with similar sampling frequency by the EROS collaboration
(\cite{afonso1998}) from Chile.  A combined analysis of this event
involving data from the MACHO/GMAN, EROS, PLANET, OGLE and MPS
collaborations was published in (\cite{afonso2000}).

Several other caustic crossings have been studied with similar
sampling frequency, by the MACHO/GMAN and PLANET collaborations
(\cite{alcock2000}; \cite{albrow2000}).  Roughly 7\% of all
microlensing events are caustic crossing events (\cite{udalski2000}).

Near a fold caustic, the magnification of a single point is
(\cite{bible})
\begin{equation}
A=A_0 + \Theta(-u_\perp) (u_\perp/u_r)^{1/2}
\end{equation}
where $u_\perp$ is the distance of the source to the caustic normal to
the caustic in units of the projected Einstein radius, $u_r$ is the
length scale of the caustic in units of the projected Einstein radius,
$A_0$ is the magnification not associated with the caustic, and
$\Theta$ is a step function.  For
typical binary lenses, $u_r$ and $A_0$ are ${\cal O}(1)$.  The
magnification of a uniform disk by a fold caustic is discussed in
(\cite{sandw1987}).  The maximum magnification is $1.4
(u_r/\rho_*)^{-1/2}$, where $\rho_*$ is the radius of the stellar disk in
units of the projected Einstein radius, while the average
magnification during the crossing is $(u_0/\rho_*)^{1/2}$.  Note that
since the planet is smaller than the source star, it is more highly
magnified.

\section{Scaling Relations}

The fraction of light reflected by a planet is
\begin{equation}
f \simeq 10^{-4} p \left ( \frac{R_p}{R_{\rm Jup}} \right )^2
	\left ( \frac{0.05 AU}{a} \right )^2 \phi
\end{equation}
where $p$ is the albedo, $R_p$ is the radius of the planet, $a$ is the
distance of the planet from the star, and $\phi$ is the fraction of the
illuminated surface of the planet visible by the observer, which depends
on the inclination and phase of the orbit.

For a source located in the Galactic bulge at $D_S = 8~\kpc$ and a
binary lens at a distance of $D_L=6~\kpc$, the typical magnification
of a planet of radius $R_p$ as it crosses the caustic is
\begin{equation}
A= 10^2 \, \left ( \frac{R_{\rm Jup}}{R_p} \right )^{1/2} \left ( \frac{M_{\rm lens}}{1 \msun} \right )^{1/2}
\end{equation}
so that, during the caustic crossing, the fractional deviation of the
light curve due to the planet is
\begin{equation}
\delta_{p} \simeq 1\% \, p \left ( \frac{R_p}{R_{\rm Jup}} \right )^{1.5}
	\left ( \frac{0.05 AU}{a} \right )^2 \left ( \frac{M_{\rm lens}}{1 \msun} \right ) ^{1/2} \phi.
\end{equation}
The duration of the planetary caustic crossing will be quite brief.
The typical planet diameter crossing time scale is
\begin{equation}
t_p=12~ \frac{R_p}{R_{\rm Jup}} \frac{100~\kms}{v_\perp} {\rm minutes}
\end{equation}
where $v_\perp$ is the component of the velocity of the planet normal
to the caustic and the line of sight.  This velocity is made up of two components, the
velocity of the planet around its star, and the velocity of the star
relative to the caustic.  In addition, the planet will not be
uniformly illuminated and thus the width of crossing will depend on
the star/planet geometry at the time of the crossing.
For example, the caustic crossing from a ``quarter-moon'' planet with the terminator
parallel to the caustic  will have an effective width 
that is two times smaller than that arising from a
``quarter-moon'' planet with the terminator perpendicular to the caustic.

Thus, we see that direct detection of a giant close-in planet requires
relative photometry accurate to better than 1\% and a sampling
frequency of minutes.
The Poisson noise for $I = 19.7$ (a G0V star at 8~kpc with
1 mag of extinction), assuming an overall throughput of $0.1$ is
\begin{equation}
\sigma \simeq 0.5\% 
{\left({t_{\rm exp} \over 1~{\rm min.}}\right)}^{-1/2} 
{\left({D \over 10~{\rm m}}\right)}^{-1},
\end{equation} 
where $t_{\rm exp}$ is the total exposure time and $D$ is the diameter
of the telescope.  Therefore, for $t_p=t_{\rm exp}=12~{\rm min.}$, the
planet will be detected with a signal-to-noise of $\delta/\sigma
\approx 6$ for a 10 meter telescope.

In addition to the caustic crossing, one may be able detect the planet
in the pre-crossing phase.  Ultimately, the magnification of the
planet inside the crossing depends on the geometry of the lens, but
can reach a factor of typically around $10$.  In this case, before the planet
crosses the caustic, the source will be approximately 0.1\% brighter
than after the caustic crossing.  This brightening lasts much longer
than the caustic crossing, $\Delta t \approx 20~{\rm hours}\,
(a_\perp/0.05 AU) \, (100\kms/v_\perp)$ where $a_\perp$ is the
planet-star distance normal to the caustic.  Since
of order 50 times more measurements will be taken of this deviation
than of the caustic, but the deviation is of order 10 times smaller,
the pre-caustic component will have a signal-to-noise of roughly half
that of the planetary caustic crossing.

This level of photometric precision and temporal sampling will
challenge present day microlensing searches, but may be possible with future microlensing follow-up searches.

\section{Sample Light curves}

In the previous section, we derived simple scaling relations to
estimate the magnitude of the deviation one might expect during a
caustic-crossing event from a close-in giant planet orbiting the
source.  We now present a few sample light curves to confirm our
estimates from \S\ 2 and illustrate the effect.

\centerline{{\vbox{\epsfxsize=9.0cm\epsfbox{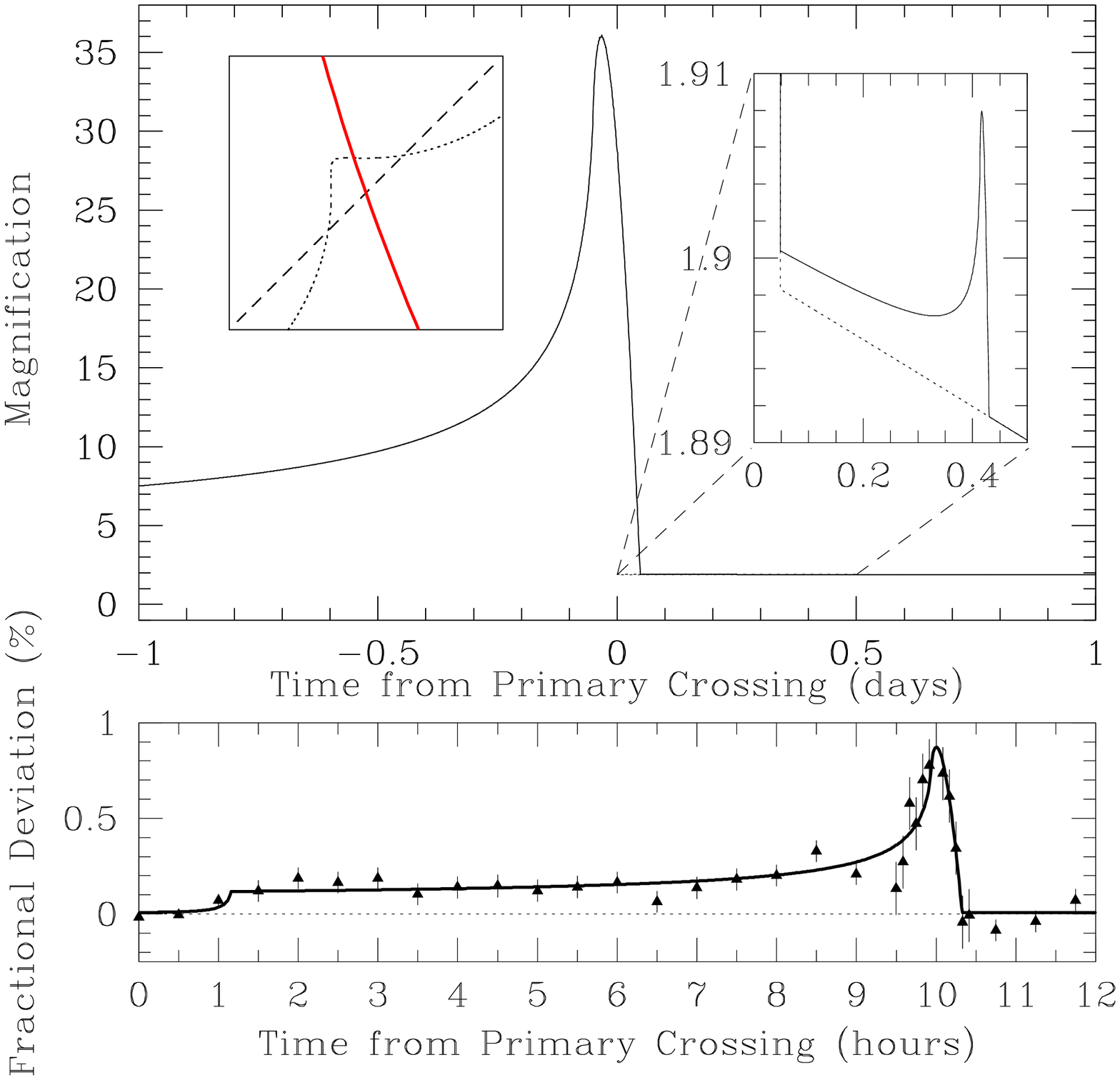}}}}
{\footnotesize {\bf FIG. 1}
Top panel: Magnification as a function of time from the primary
caustic crossing.  The lens is an equal-mass binary at 6~kpc with total mass
$1~M_\odot$.  The
source system is an analog of HD 209458 at 8~kpc:  a G0V primary with
a planet of radius $1.27~R_{\rm Jup}$ and semi-major axis
$a=0.0467~{\rm AU}$.   The top-left inset shows the geometry of the
source system trajectory:  the dashed line shows the path of the
primary while the dotted line is the path of the secondary.  The 
solid line is the caustic.  The right inset shows a blow-up of the
light curve immediately after the primary caustic crossing 
both with (solid curve) and without (dashed curve) the planet.
Bottom panel: The fractional deviation from the single-source (i.e. no
planet) light curve as a function of time from the
primary caustic crossing. The points with error bars are simulated
photometric measurements for a $I\simeq 20~{\rm mag}$ star 
assuming 5~minute exposures on a 10~m telescope and photon noise limited
precision.  Data outside the planet caustic crossing has been binned in 30 min intervals.
In this case, the deviation would be detected with a signal-to-noise
of $20$. }
%\label{fig:fig1}
%\end{figure*}
\bigskip

We adopt typical lens parameters for events toward the Galactic bulge:
we assume the lens is an equal-mass binary at 6~kpc with total mass
$1~M_\odot$, projected separation of $1~R_{\rm E}$, and transverse
velocity $v=150~\kms$.  We begin by assuming the source system is an
analog of HD 209458 at 8~kpc: a G0V primary with a planet of radius
$1.27~R_{\rm Jup}$ and semi-major axis $a=0.0467~{\rm AU}$
(\cite{charbon2000}) and the planet orbit plane is face on to the
observer.  The resulting light curve is shown in Figure 1.  Confirming
our estimates from \S\ 2, we find that the fractional deviation is
$\sim 0.1\%$ for the first $\sim 9$~hours after the primary caustic
crossing, and $0.5-1.0\%$ in the $\sim 30~{\rm min.}$ interval during
which the planet is crossing the caustic.

To illustrate the diversity of features possible, in Figure 2 we show
light curves one would expect under various assumptions about the lens
and source systems.  If the planet is very close to the primary
$a\la 0.03~{\rm AU}$, relatively large $R_{p}\ga 1.5~{\rm Jup}$,
or the lens has a low mass $M_{\rm lens}\la 0.3~M_\odot$, then the
magnitude of the deviation can be quite substantial $\delta_p \ga
1\%$, and may be detectable by current microlensing follow-up surveys
(at least for bright 

\centerline{{\vbox{\epsfxsize=9.0cm\epsfbox{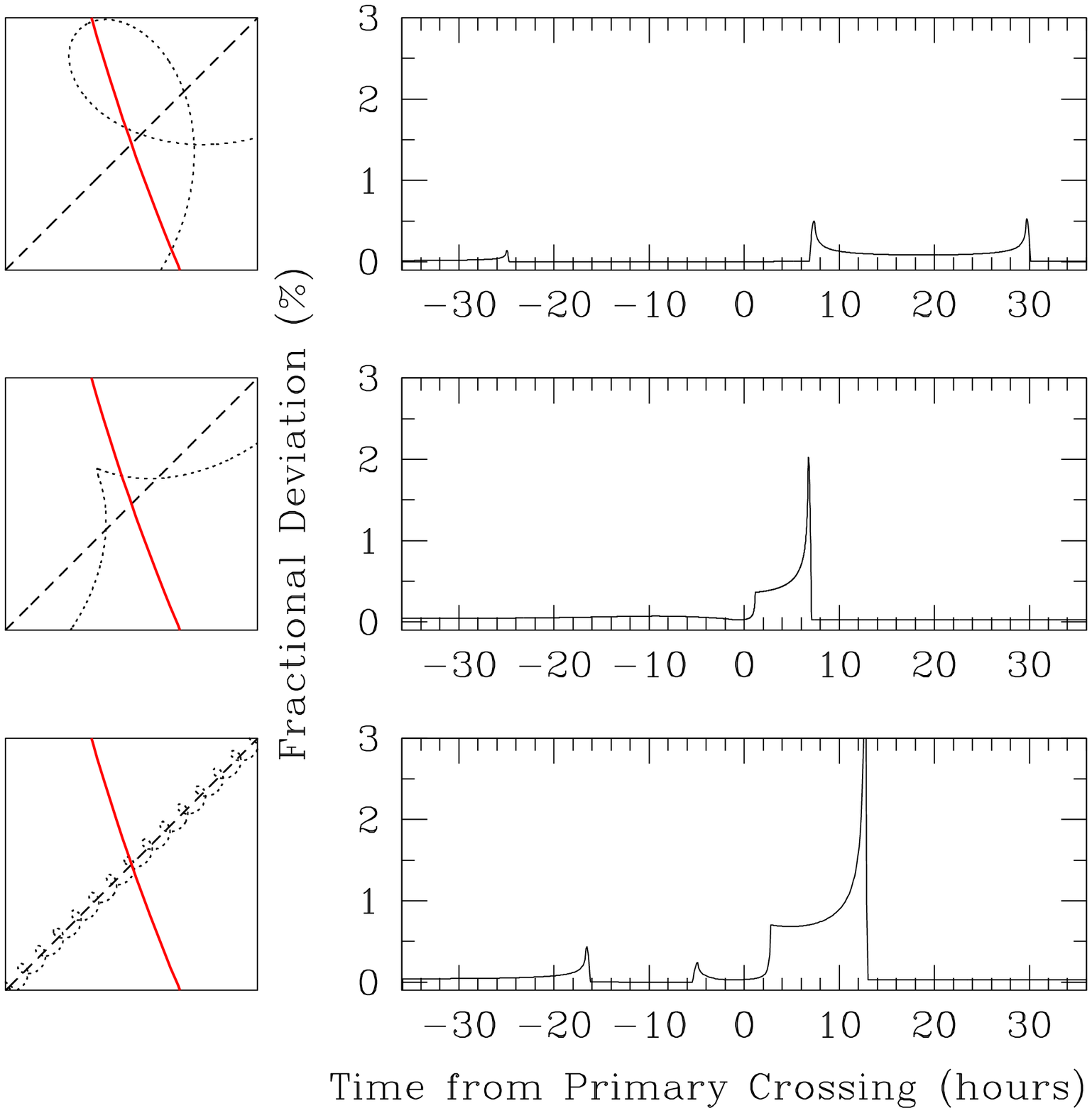}}}}
{\footnotesize {\bf FIG. 2}
%\begin{figure*}[t]
%\epsscale{0.7}
%\plotone{fig2.eps}
%\caption{ \footnotesize 
The right panels show the fractional deviation
from the single-source (i.e. no planet) light curve as a function of
time from the primary caustic crossing for three different assumptions
of the lens and/or planet parameters.  In all cases, the primary is a
G0V star at $8~{\rm kpc}$. For each panel, the geometry of the source
system trajectory is shown to the left.  Line types are the same as in
Fig.\ 1.  Top panels: Lens mass $M_{\rm lens}=0.3~M_\odot$, lens
distance $D_L=7~{\rm kpc}$, lens transverse velocity $v=60~\kms$.  The
planet has a radius $R_p=1.27~R_{\rm Jup}$ with semi-major axis
$a=0.0467~{\rm AU}$ (as in Fig.~1). Middle panels: $M_{\rm
lens}=0.3~M_\odot$, $D_L=6~{\rm kpc}$, $v=150~\kms$, $R_p=1.5~R_{\rm
Jup}$, $a=0.03~{\rm AU}$.  Bottom panels: $M_{\rm lens}=1.0~M_\odot$,
$D_L=2~{\rm kpc}$, $v=20~\kms$, $R_p=1.5~R_{\rm Jup}$, $a=0.03~{\rm
AU}$.  }
%\label{fig:fig2}
%\end{figure*}
\bigskip

\noindent sources). Also, if the period of the planetary
orbit is smaller than the time it takes for diameter of the orbit to
traverse the caustic, then the planet will cross the caustic three (or
more) times.  As we discuss in \S\ 4, this would allow one to extract
additional information about the planet.

\section{Discussion}

The current microlensing follow-ups are only marginally able to detect
planets around bright source stars by this method.  They will only detect the closest
in, largest, slowest moving giant planets around relatively bright
sources, with low signal to noise and only a handful of photometric
measurements.  However, these follow-up experiments currently
use only 1 meter class ground based telescopes and are further hindered by
their relatively poor seeing in extremely crowded fields.  A
follow-up program involving 1 night on a 10m class telescope with
adaptive optics should be able to provide millimagnitude photometry of
19th magnitude bulge sources with 5 minute time resolution, and have a
good chance of finding a close in giant planet around the source if
one is there to be found.

How many planets per year could be detected?  Currently, $\sim 50$
events are alerted toward the Galactic bulge each year; future
upgrades may provide as many as $\sim 500$ alerts per year.  Caustic
crossing binaries comprise $\sim 7\%$ of all events
(\cite{udalski2000}).  Assuming that the fraction of stars in the
bulge with close planets is similar to the local neighborhood, $\sim
1\%$, than we would expect to detect of order 1 planet per year.  This
would require a total $\sim$400 hours per season on 10m-class
telescopes.
 
Although the simple detection of a planet around a bulge star is interesting,
ultimately, one would like to measure physical parameters of the planet: its
radius, semi-major axis, and albedo.  Unfortunately, the observables
are complicated functions of not only these parameters, but also the
inclination and phase of the orbit and the mass and distance
to the lensing system.   For example,
the time scale of the caustic crossing depends on not only the
radius of the planet, but also on the projected velocity of the planet
around its star and the phase of the planet (a crescent planet will
appear to have a smaller radius than a full planet). 

Given excellent data on one caustic crossing, it would be possible to
determine the effect of the phase of the planet on the width of the
caustic crossing from the shape of the crossing alone. 
The required level of data will probably not be
possible with a 10-meter class telescope, but should be possible with a
100-meter telescope such as the OWL (\cite{owl}).
Even then, it still will not be possible to recover the velocity of the
planet, and therefore impossible to measure its radius.
However, if the transverse velocity of the star normal to the caustic is
smaller than the velocity of the planet, it is possible that there
will be three caustic crossings.  If all three are detected, each with
its own time and its own time scale, then the orbit can be solved, and
the radius determined, even without inferring the planetary phase from 
small fluctuations in the light curve.

Irrespective of whether or not the orbit can be solved, planetary
caustic crossings act like a telescope with tremendous angular
resolution perpendicular to the caustic.  The angular dependence of
the planetary albedo may be
observable. In the future, enormous telescopes such as the OWL
may be able to achieve enough temporal resolution to see the
bands and red spot on a Jupiter around a star clear across the Galaxy.

\section{Conclusions}

We have shown that the presence of a giant close-in planet around the
source star of a caustic crossing microlensing event could generate an
order 1\% deviation in the light curve.  This deviation would be
detectable with a 10 meter class telescope making measurements every
few minutes over the course of a night.  In some cases, the planet
will transit the caustic more than once, allowing for complete
solution of the orbit and a determination of the radius and albedo of the planet.
Follow-up with 100 meter class telescopes may allow for resolution of
limb darkening and spots or bands on the surface of these planets.

\smallskip
This paper profitted from useful discussions with Pierre Vermaak,
Andrew Gould and Penny Sackett.  This work was supported in part by grant AST 97-27520
from the NSF. B.S.G. acknowledges
the support of a Presidential Fellowship from the Ohio State University.

\end{document}